\documentclass[conference]{IEEEtran}
\IEEEoverridecommandlockouts

\usepackage{cite}
\usepackage{amsmath,amssymb,amsfonts}
\usepackage{algorithmic}
\usepackage{graphicx}
\usepackage{stfloats}
\usepackage{textcomp}
\usepackage[bookmarks=false]{hyperref}
\hypersetup{
    colorlinks = true,
    citecolor  = blue,
    linkcolor  = blue,
    urlcolor   = blue,
}
\usepackage{colortbl}
\usepackage[table]{xcolor}
\usepackage{multirow}
\usepackage{booktabs}
\usepackage{cleveref}
\usepackage{subcaption}
\usepackage{pifont}
\def\equationautorefname~#1\null{Eq.~(#1)\null}
\def\figureautorefname~#1\null{Fig.~#1\null}
\def\tableautorefname~#1\null{Tab.~#1\null}
\def\definitionautorefname~#1\null{Def.~#1\null}
\def\sectionautorefname~#1\null{Sec.~#1\null}
\def\subsectionautorefname~#1\null{Sec.~#1\null}
\def\subsubsectionautorefname~#1\null{Sec.~#1\null}

\def\BibTeX{{\rm B\kern-.05em{\sc i\kern-.025em b}\kern-.08em
    T\kern-.1667em\lower.7ex\hbox{E}\kern-.125emX}}

\begin{document}

\title{\raisebox{-7pt}{\includegraphics[height=29pt]{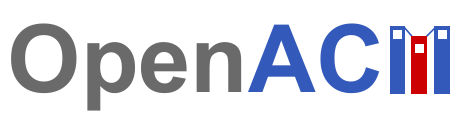}}: An {Open}-Source SRAM-Based Approximate CiM Compiler \\
\thanks{
This work is supported by the Fundamental Research Funds for the Central Universities under grant No.30924012004 and No.30925010605, and by the National Key Laboratory of Integrated Circuits and Microsystems under grant No. JCYQ2310803-1.

$^*$ Corresponding authors.
\vspace{-20pt}
}
}

\author{
\IEEEauthorblockN{
Yiqi Zhou$^{1}$,
JunHao Ma$^{1}$,
Xingyang Li$^{2}$,
Yule Sheng$^{1}$,
Yue Yuan$^{1}$,
Yikai Wang$^{1}$,
Bochang Wang$^{1}$,
Yiheng Wu$^{1}$,\\
Shan Shen$^{1,*}$,
Wei Xing$^{3,*}$,
Daying Sun$^{1,*}$,
Li Li$^{1}$ and Zhiqiang Xiao$^{4}$
}

\IEEEauthorblockA{
$^{1}$\textit{Nanjing University of Science and Technology, Nanjing, 210094, China} \\
$^{2}$\textit{Beihang University, Beijing, 100191, China} \\
$^{3}$\textit{University of Sheffield, Sheffield, S10 2TN, United Kingdom} \\
$^{4}$\textit{The 58th Research Institute of China Electronics Technology Group Corporation, Wuxi, 214035, China}
}
\vspace{-20pt}
}

\maketitle

\begin{abstract}
The rise of data-intensive AI workloads has exacerbated the ``memory wall'' bottleneck. Digital Compute-in-Memory (DCiM) using SRAM offers a scalable solution, but its vast design space makes manual design impractical, creating a need for automated compilers. A key opportunity lies in approximate computing, which leverages the error tolerance of AI applications for significant energy savings. However, existing DCiM compilers focus on exact arithmetic, failing to exploit this optimization.
This paper introduces OpenACM, the first open-source, accuracy-aware compiler for SRAM-based approximate DCiM architectures. OpenACM bridges the gap between application error tolerance and hardware automation. Its key contribution is an integrated library of accuracy-configurable multipliers (exact, tunable approximate, and logarithmic), enabling designers to make fine-grained accuracy-energy trade-offs. 
The compiler automates the generation of the DCiM architecture, integrating a transistor-level customizable SRAM macro with variation-aware characterization into a complete, open-source physical design flow based on OpenROAD and the FreePDK45 library. 
This ensures full reproducibility and accessibility, removing dependencies on proprietary tools. 
Experimental results on representative convolutional neural networks (CNNs) demonstrate that OpenACM achieves energy savings of up to 64\% with negligible loss in application accuracy. The framework is available on \href{https://github.com/ShenShan123/OpenACM}{OpenACM:URL}.
\end{abstract}


\section{Introduction}


The growth of data-intensive Artificial Intelligence (AI) applications has intensified the ``memory wall" bottleneck in traditional von Neumann architectures. The constant data transfer between memory and processing units severely limits both performance and energy efficiency~\cite{hardware_approximate_techniques}. Compute-in-Memory (CiM) directly addresses this challenge by performing computations within the memory array, thereby minimizing data movement.


Among CiM approaches, Digital Compute-in-Memory (DCiM), which integrates logic into standard SRAM arrays, has emerged as a scalable and practical solution~\cite{openc2_2025, sram_cim_review}. However, the architectural flexibility of DCiM creates a vast design space spanning choices in bit-cell design, arithmetic circuits, and array organization. Manually exploring this space is prohibitively slow, rendering comprehensive Design Space Exploration (DSE) impractical for the rapid hardware iteration that AI applications demand. This challenge necessitates sophisticated compilers to automate the design flow from high-level specification to physical layout.


Concurrently, approximate computing provides a powerful technique for improving energy efficiency. Many AI applications possess an inherent tolerance to minor computational errors~\cite{exploiting_approximate_computing, approxann_framework}. By strategically introducing controlled approximations into the hardware, designers can trade negligible losses in application-level accuracy for significant reductions in power consumption and area.

Existing DCiM compilers primarily focus on \textit{exact} arithmetic~\cite{autodcim2023, arctic2024, syndcim2024, segadcim2025, damildcim2025} and lack support for approximation, thereby failing to exploit the full energy-saving potential permitted by error-tolerant applications. Moreover, current tools do not provide precision-configurable DCiM architectures that can adapt to varying levels of application-specific error tolerance. The reliance of many compilers on proprietary EDA toolchains further limits reproducibility and inhibits broader community-driven development~\cite{anysilicon_eda_guide}.

To address these limitations, this work presents \textbf{OpenACM}, an \underline{Open}-source SRAM-based \underline{A}pproximate \underline{C}i\underline{M} compiler. OpenACM is the first open-source framework that systematically integrates accuracy-aware approximate computing into a fully automated DCiM design flow. Enabled by flexible bit-width and precision configurability, OpenACM provides a unified platform for exploring energy-efficient CiM circuits under different application-specific accuracy constraints.  Our key contributions are:
\begin{itemize}
\item \textbf{Approximate DCiM compiler with accuracy-configurable multipliers:} 
OpenACM integrates a multiplier library offering three families selectable under application accuracy constraints: (i) an exact 4-2 compressor-based multiplier; (ii) an approximate 4-2 compressor-based multiplier with tunable accuracy; and (iii) a logarithmic approximate multiplier. This enables fine-grained accuracy-energy trade-offs within a unified flow.

\item \textbf{Complete open-source ecosystem integration:} The back-end (physical) design flow is implemented using the OpenROAD digital design flow~\cite{openroad} with FreePDK45~\cite{freepdk45} to ensure full reproducibility without proprietary dependencies. The SRAM macro follows a FakeRAM2.0-style~\cite{fakeram2} template for seamless macro abstraction and Physical Design (PD) integration~\cite{fakeram2}.

\item \textbf{Variation-Aware SRAM Analysis:} Built upon OpenYield~\cite{openyield}, OpenACM supports transistor-level customization of SRAM macros and integrates importance-sampling-based Monte Carlo (MC) simulation to accelerate library characterization under Process Voltage Temperature (PVT) variations, together with automated transistor sizing optimizations.
\end{itemize}

The remainder of this paper is organized as follows: \autoref{sec:background} reviews relevant background and related DCiM compilers; \autoref{sec:architecture} presents the OpenACM framework and its core components; \autoref{sec:flow} details the OpenROAD-based physical design flow; \autoref{sec:results} reports post-layout results; and \autoref{sec:conclusion} concludes with discussions and future directions.
OpenACM is under continuous development, with several key future enhancements discussed in \autoref{sec:conclusion}.

\section{Background}
\label{sec:background}
\subsection{Error-Tolerant Applications}
The fundamental premise of this work rests on a powerful opportunity: the inherent error resilience of neural networks. Unlike traditional high-performance computing, DNNs are statistical models that learn to recognize patterns in noisy, real-world data. Their robustness is analogous to the human brain's ability to identify a familiar face in a poorly lit photograph; perfect, high-fidelity input is not required for a correct outcome. This resilience translates directly to the hardware level. The millions of multiply-accumulate operations underpinning a network's inference do not all require perfect mathematical accuracy. A small degree of computational error in circuits can often be introduced with little to no degradation in the final application accuracy. This presents a golden opportunity to trade this unneeded accuracy for dramatic savings in energy consumption and silicon area~\cite{venkataramani2015approximate, mrazek2022hardware}.

\subsection{Digital CiM Compilers}
Automated DCiM compilers~\cite{autodcim2023, arctic2024, syndcim2024, segadcim2025, damildcim2025, openc2_2025} have significantly improved design productivity. However, none currently support configurable approximation or comprehensive yield-aware SRAM analysis. \autoref{tab:dcim_comparison} highlights the key distinctions in precision, openness, and analysis depth.
\begin{table}[b]
\centering
\caption{The Landscape of Automated DCiM Compilers: A Missing Capability.}
\label{tab:dcim_comparison}
\setlength{\tabcolsep}{2pt}
\resizebox{\columnwidth}{!}{%
\begin{tabular}{@{} p{2.4cm} p{1.8cm} c p{1.4cm} p{2.4cm} @{} }
\toprule
\textbf{Compiler} & \textbf{Precision} & \textbf{Open?} & \textbf{SRAM Anal.} & \textbf{Multiplier}\\ 
\midrule
AutoDCIM~\cite{autodcim2023} & Fixed-point & \textcolor{red}{\ding{55}} & Basic & Adder-tree \\ 
ARCTIC~\cite{arctic2024}  & Int/FP with BIST & \textcolor{red}{\ding{55}} & Limited & Adder-tree \\
SynDCIM~\cite{syndcim2024}  & Multiple precision & \textcolor{red}{\ding{55}} & Basic & Adder-tree with 4-2 compressor \\
SEGA-DCIM~\cite{segadcim2025}  & Int/FP precision & \textcolor{red}{\ding{55}} &  Limited & Adder-tree \\
DAMIL-DCIM~\cite{damildcim2025} & Digital precision & \textcolor{red}{\ding{55}} & Basic & Adder-tree \\
OpenC\textsuperscript{2}~\cite{openc2_2025} & Digital precision & \textcolor{blue}{\ding{51}} & Basic & Adder-tree \\
\rowcolor{green!15}
OpenACM (Ours) & Configurable approximation & \textcolor{blue}{\textbf{\ding{51}}} & Advanced & Exact / Approx 4-2 compressor / Log \\
\bottomrule
\end{tabular}
}
\end{table}
As detailed in the table, AutoDCIM~\cite{autodcim2023} set the precedent for automation but is closed-source. Subsequent works like ARCTIC~\cite{arctic2024}, SynDCIM~\cite{syndcim2024}, SEGA-DCIM~\cite{segadcim2025}, and DAMIL-DCIM~\cite{damildcim2025} advanced specific aspects such as multi-precision support and physical layout, yet they remain proprietary and restricted to exact arithmetic. OpenC\textsuperscript{2}~\cite{openc2_2025} broke the open-source barrier but offers only basic SRAM analysis and lacks approximation features.

OpenACM distinguishes itself as the first open-source compiler to systematically integrate accuracy-configurable approximation with advanced variation-aware SRAM analysis. By bridging the gap between exact-computing tools and the error tolerance of AI applications, it provides a comprehensive solution for next-generation energy-efficient designs.

\section{\raisebox{-3pt}{\includegraphics[height=12pt]{figs/openACM-logo.pdf}}: Approximate DCiM Compiler}
\label{sec:architecture}

This section overviews OpenACM's flow, beginning with the overall architecture and followed by compiler components.

\subsection{Overall Architecture}

\begin{figure}[tbp]
	\centering
	\includegraphics[width=0.9\linewidth]{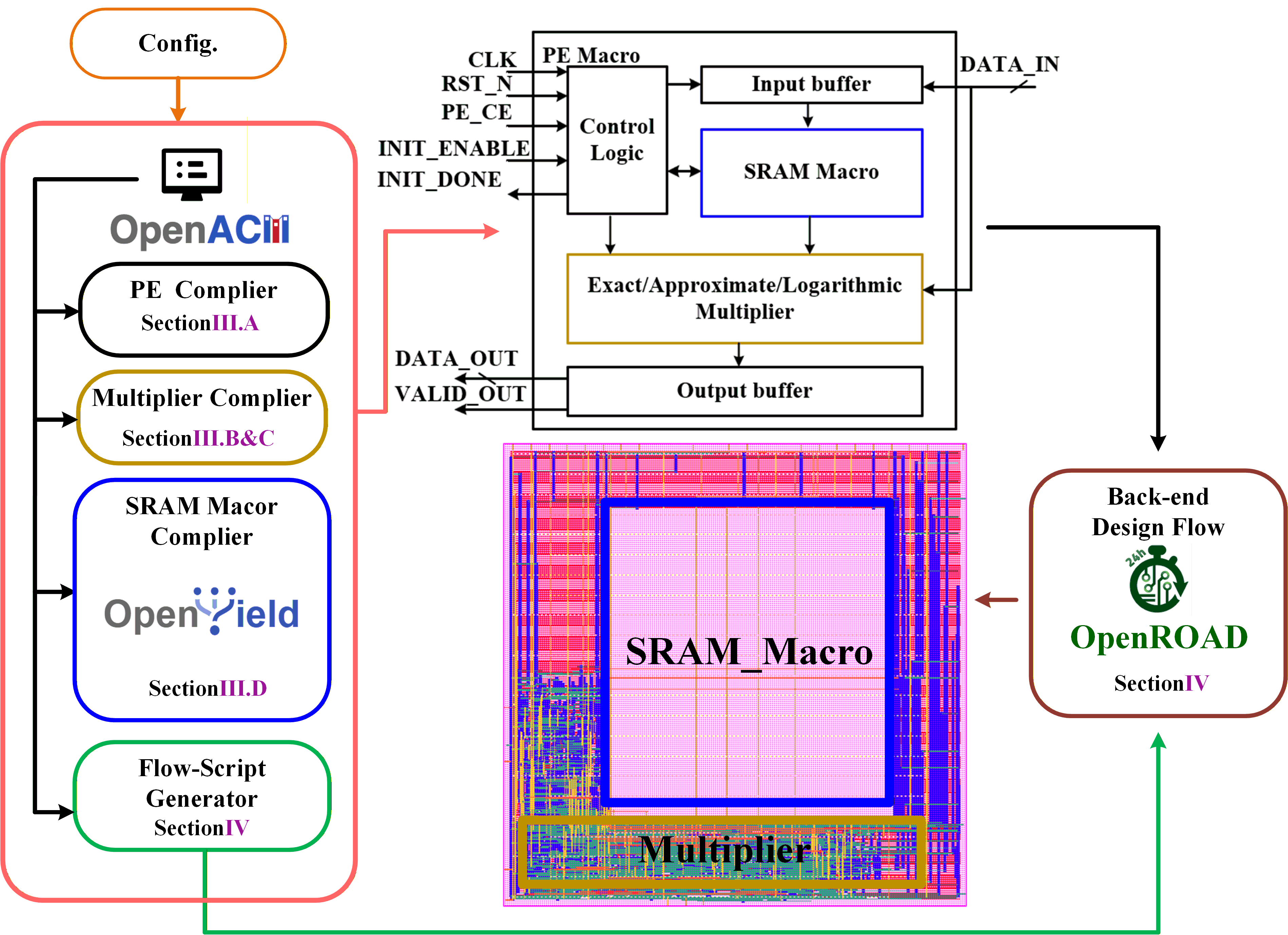}
	\caption{The end-to-end open-source workflow enabled by OpenACM, comprised of its core hardware components and the EDA toolchain. The flow integrates a library of accuracy-configurable multipliers, a generator for a customizable SRAM macro, and a script generator for the steps in the PD.}\label{Overall Architecture}
\end{figure}

As illustrated in \autoref{Overall Architecture}, OpenACM is an end-to-end framework that generates a DCiM macro from architecture specifications and multiplier configurations, surpassing traditional DCiM compilers that rely solely on bit-width scaling for accuracy control.

The OpenACM compiler consists of four main components:
\begin{enumerate}
\item \textbf{Processing Element (PE) compiler} generates the control logic for the SRAM and multiplier, together with the associated input/output buffers. The PE first initializes the SRAM with the required data, and then performs multiplication between incoming data and the stored values, producing the final results.
\item \textbf{Multiplier compiler} provides (i) exact multipliers of arbitrary bit widths; (ii) approximate multipliers with configurable precision using approximate 4-2 compressors to optimize the partial-product reduction tree; and (iii) logarithmic multipliers to further improve energy efficiency, especially for large bit-width designs. 
\item \textbf{SRAM macro compiler} generates 6T SRAM arrays of arbitrary dimensions, along with the necessary control and read/write circuitry.
\item \textbf{Flow-script generator} produces the backend design scripts required by OpenROAD and leverages it to complete the PD flow. The compiler ultimately delivers a physically implemented CiM macro together with sign-off artifacts, enabling designers to meet application-specific accuracy and energy-efficiency requirements without relying on proprietary layout generation tools.
\end{enumerate}


\subsection{Approximate Multiplier}\label{sec:app}

Existing DCiM compilers mainly use fixed-point representations with varying bit widths, offering only coarse-grained accuracy control. However, diverse application error tolerances demand finer tuning. To address this, we introduce approximate computing techniques that leverage approximate compressors under a given bit width for more precise accuracy adjustment. In SynDCIM~\cite{syndcim2024}, the use of exact 4-2 compressors was proposed to optimize adder trees, but their effectiveness remained constrained by the strict requirement of exact computation. In contrast, OpenACM integrates approximate multipliers based on approximate 4-2 compressors, enabling flexible trade-offs between accuracy and PPA, and significantly expanding the design space of DCiM beyond exact computing.
\begin{figure}[tb]
	\centering
	\includegraphics[width=\linewidth]{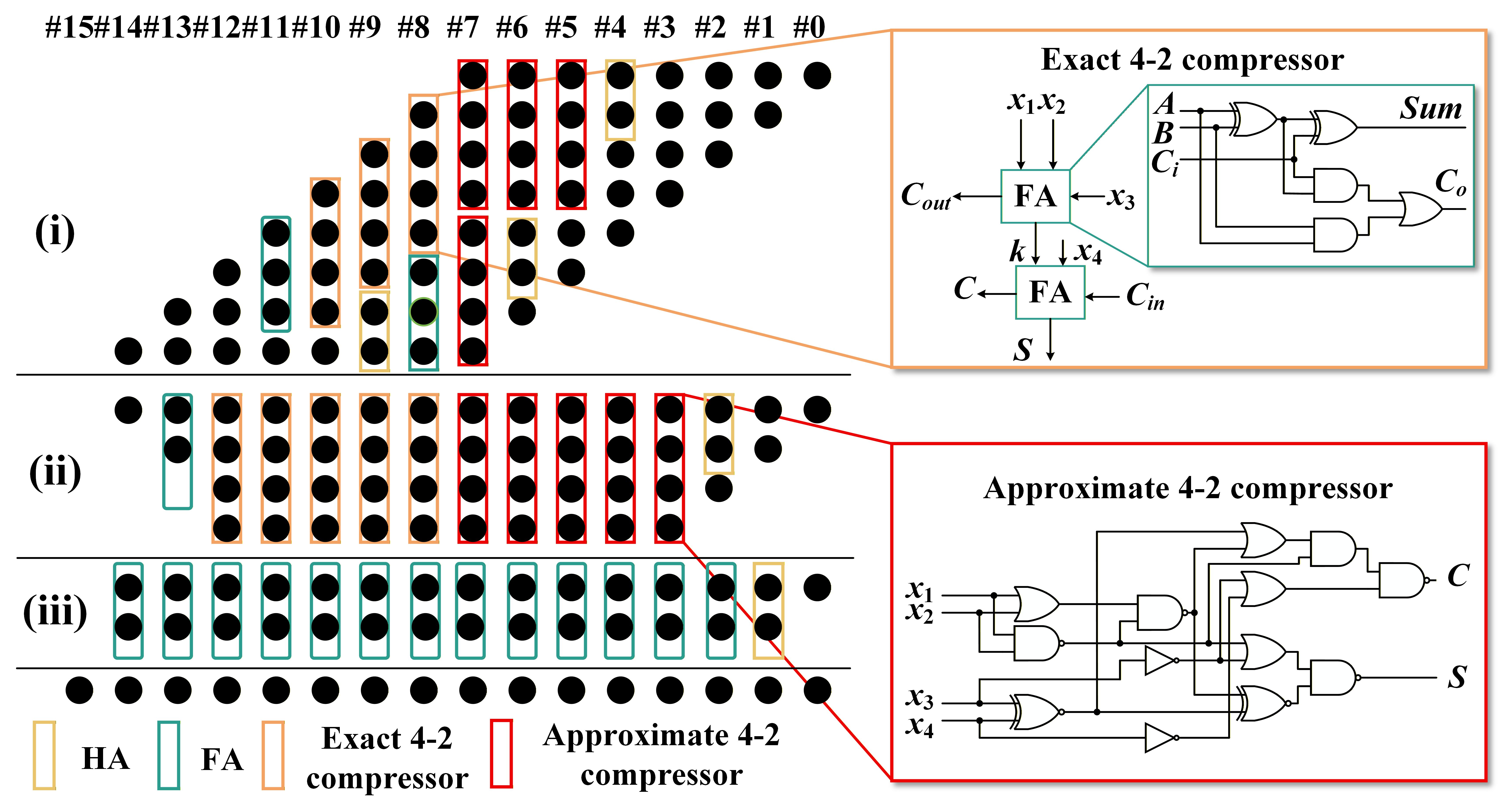}
	\caption{Block diagram of the proposed 8-bit approximate multiplier. It consists of three stages: (i) partial-product generation, (ii) a configurable reduction tree employing exact or approximate 4-2 compressors on selected low-order columns, and (iii) a final carry-propagate adder.}\label{Figure_2}
\end{figure}
For approximate multipliers based on 4-2 compressors, the circuit structure is typically depicted in \autoref{Figure_2}, which presents an 8-bit multiplier. In the initial stage, black dots indicate Partial Products (PPs) generated by AND gates from the two input operands. These PPs are compressed through two intermediate stages, ultimately resulting in two rows of PPs in the final stage. These rows are then summed to produce the output of the multiplier. Throughout the three stages, various combinational logic circuits, including HAs, FAs, and 4-2 compressors, are employed to compress the PPs. To optimize resource consumption while maintaining acceptable accuracy, approximate 4-2 compressors are commonly applied in the lower 8 bits of the PPs, specifically in columns \#0 to \#7, as highlighted by the red box in \autoref{Figure_2}. The framework is fully scalable and supports approximate multipliers of arbitrary bit widths. Designers can either tailor approximate 4-2 compressors to meet specific accuracy requirements or adopt widely-used approximate 4-2 designs~\cite{akbari2017dual,ha2017multipliers,kong2021design,momeni2014design,yang2015approximate,strollo2020comparison}. Moreover, OpenACM also supports the generation of exact multipliers of any bit width, enabling flexible selection of multipliers that best satisfy the accuracy requirements of different applications. 

\subsection{Logarithmic Multiplier}\label{sec:log}
Although approximate 4-2 compressor multipliers improve area and power efficiency, applications with high error tolerance allow further accuracy relaxation. To extend the design space beyond compressor-based approximation, OpenACM integrates an energy-efficient Logarithmic Multiplier (LM), which naturally provides finer-grained accuracy-PPA trade-offs.

The conventional LM is derived from Mitchell’s Algorithm (MA)~\cite{mitchell2009computer}, which approximates multiplication in the logarithmic domain. For an operand $N$, it can be expressed as $N = 2^k(1 + x)$, where $k$ denotes the position of the most significant ``1'' and $x$ represents the fractional part. This expression can also be reformulated as $x \cdot 2^k = N - 2^k$. Consequently, the product of two operands $A$ and $B$ can be written as:
\begin{equation}
\begin{split}
A \times B = & {2^{{k_1} + {k_2}}} + (A - {2^{{k_1}}}){2^{{k_2}}} + (B - {2^{{k_2}}}){2^{{k_1}}} \\
             & + (A - {2^{{k_1}}})(B - {2^{{k_2}}})
\end{split}
\end{equation}

Where $k_1$ and $k_2$ represent the leading-one position of $A$ and $B$, respectively, the $(A - {2^{{k_1}}})(B - {2^{{k_2}}})$ is taken as the Error Part (EP) and the remainder is taken as the Approximate Part (AP). After taking the logarithm on both sides, the AP can be computed using only shift and addition operations, while the EP is typically neglected or approximated. Neglecting the EP introduces significant error; therefore, we propose an adder-free dynamic compensation strategy to handle the EP term.

For an $n$-bit multiplier, the maximum Rounding Error (RE) associated with the leading-one position $k$ can be expressed as:
\begin{equation}
    {\rm RE}_k = (2^{k+1} - 2^k)/2 = 2^{k-1}, \quad k \in \{1,2,\dots,n-2\}.
\end{equation}

The Worst-Case Error (WCE) depends on which operand is chosen for rounding in the EP, where $Q_1 = A - 2^{k_1}$ and $Q_2 = B - 2^{k_2}$. If the smaller operand is rounded, the WCE is $4^{n-2} - 2^{n-3}$; if the larger operand is rounded, the WCE reduces to $3 \cdot 4^{n-3}$. To minimize the WCE, the proposed algorithm dynamically selects and scales the larger operand in the EP to generate compensation values.

Additionally, to incorporate error compensation without hardware overhead, the proposed compensation algorithm generates compensation values within a unique range. Since operands $Q_1$ and $Q_2$ are derived from shifted values with their leading ones removed, the range of the compensation value ${\rm{round}}({Q_1}){Q_2}$ is strictly less than ${2^{{k_1} + {k_2}}}$. This guarantees that when the compensation is added to ${2^{{k_1} + {k_2}}}$, no carry is generated. Thus, a bitwise OR gate can replace an FA without affecting correctness. The final approximate product can be expressed as:
\begin{equation}
    {P_{approx}} = {2^{{k_1} + {k_2}}}|{\rm{round}}({Q_1}){Q_2} + (A - {2^{{k_1}}}){2^{{k_2}}} + (B - {2^{{k_2}}}){2^{{k_1}}}
\end{equation}

\begin{figure}[tb]
	\centering
	\includegraphics[width=\linewidth]{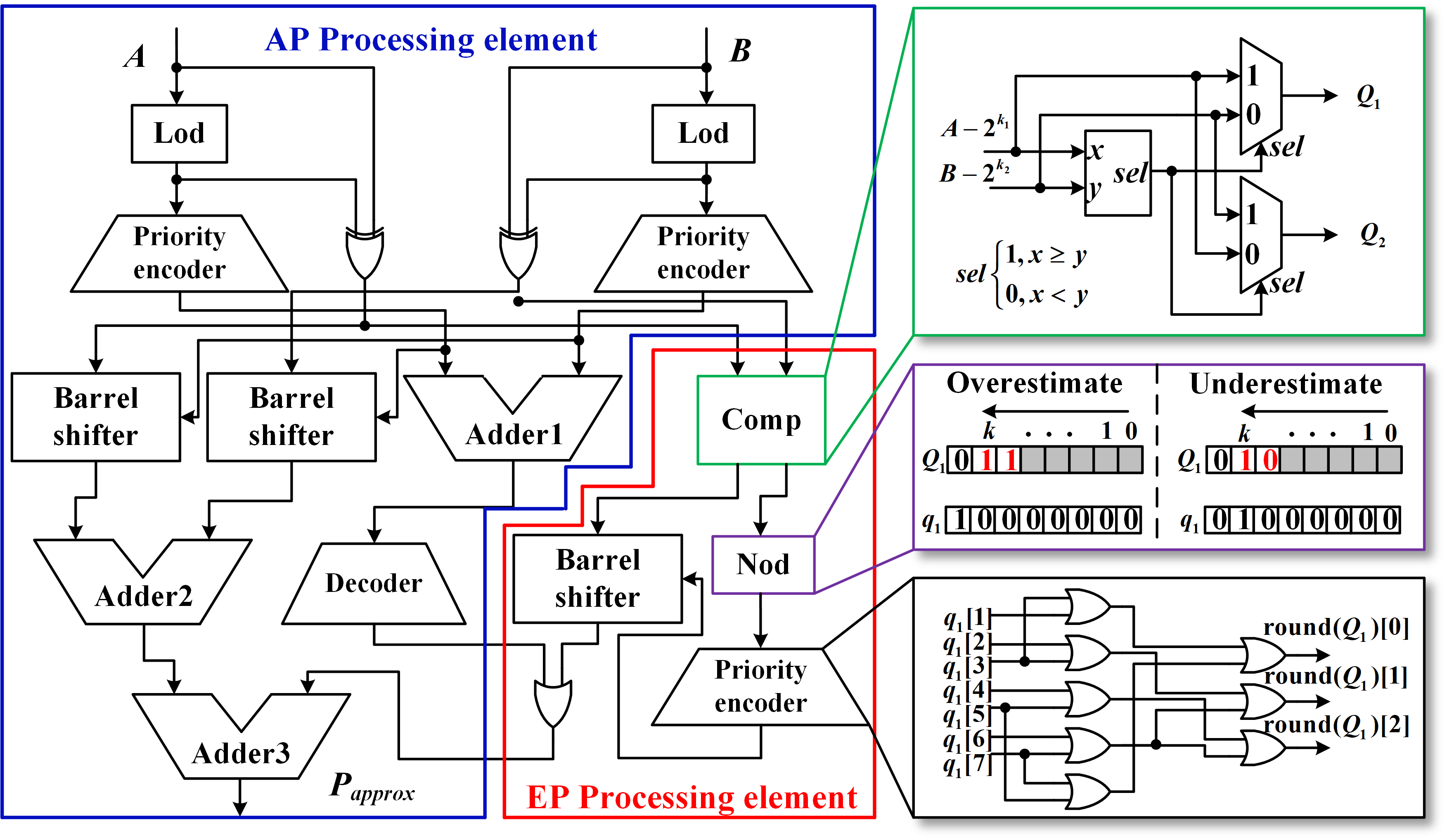}
	\caption{Block diagram of the proposed 8-bit logarithmic multiplier. It approximates multiplication by (i) base-2 logarithmic encoding using a leading-one detector and a small LUT, (ii) addition in the log domain, and (iii) antilogarithmic decoding via a barrel shifter and LUT, optionally with error-compensation.}\label{Figure_3}
\end{figure}

Finally, the proposed LM is shown in \autoref{Figure_3}. In the AP processing element, two Leading-one Detectors (LoDs) and two priority encoders are implemented to achieve the leading-one position of two operands, $k_1$ and $k_2$ represent the leading-one position of two operands, respectively. XOR gates are used to remove the leading-one from two operands, $A - {2^{{k_1}}}$ and $B - {2^{{k_2}}}$ can be obtained, $k_1$ and $k_2$ are added by the Adder1, the result of the Adder1 is decoded to the ${2^{{k_1} + {k_2}}}$. $A - {2^{{k_1}}}$ and $B - {2^{{k_2}}}$ are shifted $k_2$ and $k_1$ bits by two barrel shifters. The partial sum of $(A - {2^{{k_1}}}){2^{{k_2}}}$ and $(B - {2^{{k_2}}}){2^{{k_1}}}$ are added by the Adder2. In the EP processing element, the product of $A - {2^{{k_1}}}$ and $B - {2^{{k_2}}}$ are estimated. Two operands are compared in the COMP, and the larger is overestimated to ${2^{k + 1}}$ or underestimated to ${2^{k}}$. Then the compensated result can be obtained by shifting the smaller operand. The partial sum of the ${2^{{k_1} + {k_2}}}$ and the result of the EP processing element can be achieved by an OR gate, and intermediate values are combined by Adder3 to produce the approximate product. This architecture easily scales to different bit widths by proportionally widening each module.


\subsection{SRAM Macro}\label{sec:sram}

\begin{figure}[tb]
    \centering
    \includegraphics[width=\linewidth]{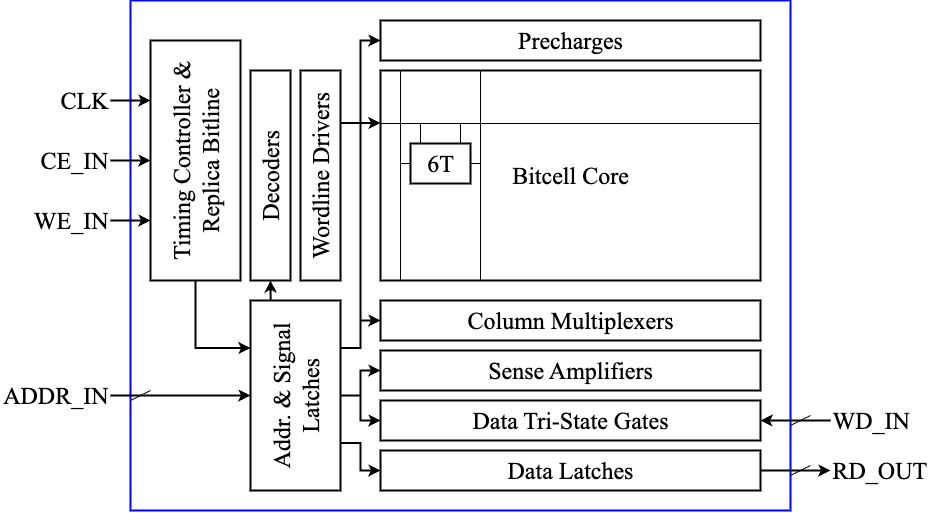}
    \caption{OpenACM SRAM macro: banked, subarrayed 6T array with hierarchical WL, PRE, write drivers, column MUX, and differential SAs.}
    \label{fig:sram_macro}
\end{figure}

The SRAM macro provides the memory substrate for DCiM operations. As shown in \autoref{fig:sram_macro}, OpenACM adopts a compact, banked, and subarrayed 6T design with hierarchical Word-Line (WL) decoders/drivers, PREcharge (PRE), write drivers, optional column multiplexers, and differential Sense Amplifiers (SAs). Reads precharge BL/BLB, assert WL, and latch via SA; writes drive BL/BLB while WL is asserted. This structure is intentionally minimal to ease timing closure and tiling across banks/subarrays. 
We highlight the following features in the implementation of the SRAM macro:
\begin{enumerate}
\item MC- and yield-aware SRAM characterization (from OpenYield~\cite{openyield}): OpenACM runs Monte Carlo SPICE with process variations (local mismatch) and built-in yield analysis algorithms to extract distributions of access latency, SNM (read/write/hold), and dynamic/leakage power; to improve efficiency, it employs the Importance Sampling (IS) method to bias rare-event regions, substantially reducing the number of simulations required and accelerating LIB characterization while preserving accuracy at the target yield. 
\item This macro supports fine-grained, compiler-visible tunable knobs for co-optimization: rows/cols, word width, bank/subarray count, column-mux ratio, and timing controls (e.g., SAE, precharge), which map directly to design-space exploration with approximate multipliers for accuracy-PPA co-design. 
\item Flow readiness and reproducibility: abstract views and footprints align with open-source flows (OpenROAD) and FakeRAM-style \cite{fakeram2} abstract macros to enable black-box integration during place-and-route; see \autoref{sec:flow} for flow details.
\end{enumerate}

Furthermore, by adhering to the FakeRAM2.0-style abstract memory interface, our SRAM macro can be seamlessly integrated into other open-source projects that already use FakeRAM macros (e.g., OpenROAD's tinyRocket tutorial~\cite{micro2022tutorial} that uses fakeram45\_256x16/fakeram7\_256x32), enabling drop-in replacement in broader SoC flows.


\section{Design Flow: OpenROAD's Toolchain}
\label{sec:flow}
\begin{figure}[tbp]
	\centering
	\includegraphics[width=0.9\linewidth]{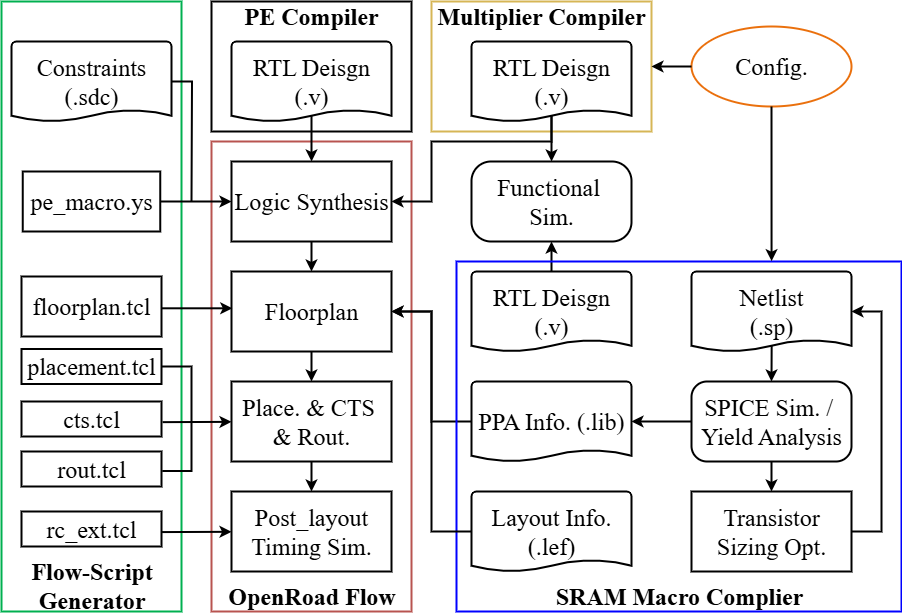}
	\caption{Overall design flow of OpenACM, which is built on top of OpenROAD and OpenYield.}\label{Figure_flow}
\end{figure}
As shown in \autoref{Figure_flow}, OpenACM accomplishes circuit implementation through a standard digital design flow. OpenACM enables automated circuit generation via Python scripts, supporting configurable parameters such as the number of SRAM rows/columns, multiplier bit-width, approximate multipliers based on 4-2 compressors, and logarithmic multipliers. For compressor-based approximate multipliers, users can further specify the compressor type and the combination strategy of different approximate compressors. Once the configuration parameters are defined, the toolchain generates: (i) an SRAM behavioral model, (ii) an LEF abstract for physical integration, (iii) LIB timing/power/area views derived from characterization, (iv) RTL for the PE control logic and multipliers, and (v) the scripts required for OpenROAD todte complete the backend physical design. Note that the current release does not generate the exact SRAM layout; the SRAM is integrated as a black-box hard macro during place-and-route. As we want to support transistor sizing customization, the GDSII generation of the SRAM macro is still under development and will be left for future work. OpenROAD implements the standard-cell logic and performs top-level integration around the SRAM abstract, with STA conducted using OpenSTA. After detailed routing, parasitic extraction (SPEF) and SDF generation are applied to the synthesized logic and interconnect (excluding SRAM internals) for post-layout timing simulation, ensuring the design meets the specified functionality and performance requirements.

\section{Experiments}
\label{sec:results}

We evaluate the effectiveness of OpenACM through a set of experiments targeting three key research questions. All CiM macros are generated using the FreePDK45~\cite{freepdk45} technology and implemented with the OpenROAD flow~\cite{openroad}, from synthesis to layout. Power and energy are derived from post-layout switching activity simulations, while SRAM behavior is characterized using Xyce~\cite{keiter2013xyce} circuit simulations.
\subsection{PPA Comparison}
\begin{table}[tbp]
\centering
\caption{Post-layout performance of OpenACM-generated SRAM-multiplier systems at 100\,MHz and a 0.5\,pF output load.}
\label{tab:sram_performance}
\resizebox{\linewidth}{!}{%
\begin{tabular}{ccccccccc}
\toprule
\multirow{2}{*}{SRAM} & \multirow{2}{*}{Multiplier Type} & \multirow{2}{*}{Delay (ns)} & \multicolumn{3}{c}{Area ($\mu$m\textsuperscript{2})} & \multirow{2}{*}{Power (W)} \\
\cmidrule{4-6}
 & & & Logic & SRAM & P\&R & \\
\midrule
\multirow{4}{*}{\begin{tabular}[c]{@{}c@{}}16 $\times$ 8\\ (8-bit width)\end{tabular}} 
 & OpenC\textsuperscript{2} & 5.22 & 1431 & \multirow{4}{*}{7052} & 8483 & 2.82E-04 \\
 & Exact & 5.22 & 1079 &  & 8131 & 2.45E-04 \\
 & Log-our & 5.22 & 1173 &  & 8225 & 2.82E-04 \\
 & \textbf{Appro4-2} & 5.22 & \textbf{939} &  & \textbf{7991} & \textbf{2.11E-04} \\
\midrule
\multirow{4}{*}{\begin{tabular}[c]{@{}c@{}}32 $\times$ 16\\ (16-bit width)\end{tabular}} 
 & OpenC\textsuperscript{2} & 5.24 & 4842 & \multirow{4}{*}{16910} & 21752 & 1.15E-03 \\
 & Exact & 5.24 & 3568 &  & 20478 & 1.08E-03 \\
 & \textbf{Log-our} & 5.24 & \textbf{2402} &  & \textbf{19312} & \textbf{6.15E-04} \\
 & Appro4-2 & 5.24 & 2633 &  & 19543 & 7.58E-04 \\
\midrule
\multirow{4}{*}{\begin{tabular}[c]{@{}c@{}}64 $\times$ 32\\ (32-bit width)\end{tabular}} 
 & OpenC\textsuperscript{2} & 5.24 & 19734 & \multirow{4}{*}{48642} & 68376 & 7.00E-03 \\
 & Exact & 5.24 & 10132 &  & 58774 & 4.03E-03 \\
 & \textbf{Log-our} & 5.24 & \textbf{4960} &  & \textbf{53602} & \textbf{1.45E-03} \\
 & Appro4-2 & 5.24 & 9331 &  & 57973 & 3.36E-03 \\
\bottomrule
\end{tabular}%
}
\end{table}

For a fair comparison, it is worth noting that TOPS/W is not reported since it mainly reflects architecture-level throughput and is heavily affected by interconnect and scheduling, making it unsuitable for PE macro-level evaluation where SRAM access and control dominate performance. Instead, we present area, delay, and power, which directly reveal the circuit efficiency of our PE and provide a more reliable basis for system-level scaling. \autoref{tab:sram_performance} presents the post-layout results of SRAM-multiplier systems generated by OpenACM under a 100 MHz clock and a 0.5 pF load. The compiler automatically assembles the SRAM arrays, control units, and three arithmetic cores: an exact multiplier using exact 4-2 compressors, an approximate design based on approximate 4-2 compressors (Appro4-2), and the proposed logarithmic multiplier (Log-our). Since OpenACM supports arbitrary combinations of approximate 4-2 compressors, we adopt the widely used and highly cited Yang1~\cite{yang2015approximate} as a representative example to demonstrate the energy-efficiency advantages of approximate CiM compiler. For completeness, we also include a representative adder-tree-based implementation from OpenC\textsuperscript{2} as a baseline. 

All designs are evaluated using the same multiplication workloads to ensure fair power comparison. The results show that the critical delay is nearly constant (5.2 ns) across all multipliers, indicating SRAM dominates system timing. Approximate designs achieve significant area reduction: the logarithmic multiplier cuts logic area by 33\% for $32 \times 16$ and 51\% for $64 \times 32$. For energy, the 4-2 compressor design saves up to 14\% power at $16 \times 8$, favoring small-scale cases, while the logarithmic design excels at larger scales, reducing power by nearly 64\% compared to the exact multiplier in $64 \times 32$ and also outperforming Appro4-2. By contrast, the adder-tree-based architecture (OpenC\textsuperscript{2} \cite{openc2_2025}) exhibits consistently higher area and power consumption across all configurations. It is worth noting that for small bit-widths (e.g., 8-bit), the logarithmic multiplier exhibits a slightly higher overhead than the 4-2-based structure. This is because the logarithm and anti-logarithm estimation modules constitute a relatively large portion of the short datapath, introducing additional logic complexity and switching activity. In contrast, the 4-2 compressor network remains structurally compact and therefore offers advantages in both area and dynamic power at low bit-widths. As bit-width increases, the relative overhead of the logarithmic modules diminishes, enabling the logarithmic multiplier to deliver substantial area and energy benefits in medium- and large-scale configurations.


\subsection{Accuracy-Constrained Applications}

\begin{table}[tbp]
\centering
\caption{PSNR Comparison of Different Approximate Multipliers for Various Image Processing Tasks.}
\label{tab:Image Processing}
\resizebox{\linewidth}{!}{%
\begin{tabular}{>{\centering}m{2cm} >{\centering}m{3cm} c c c c}
\toprule
\multirow{2}{*}{Processing Task} & \multirow{2}{*}{Test Image} & \multicolumn{3}{c}{Multiplier Type} \\
\cmidrule{3-5}
 & & Appro4-2 & Log-our & LM~\cite{mitchell2009computer} \\
\midrule
\multirow{3}{*}{Image Blending} 
    & Lake \& Mandril   & 67.19 dB & 32.01 dB  & 26.08 dB \\
    & Jetplane \& Boat  & 70.93 dB & 37.17 dB  & 22.10 dB \\
    & Cameraman \& Lake & 69.81 dB & 43.22 dB  & 24.82 dB \\
\midrule
\multirow{3}{*}{Edge Detection}
    & Boat              & 66.21 dB & 46.43 dB  & 38.77 dB \\
    & Cameraman         & 67.55 dB & 45.61 dB  & 38.37 dB \\
    & Jetplane          & 66.20 dB & 44.13 dB  & 39.07 dB \\
\bottomrule
\end{tabular}%
}
\end{table}

Image blending~\cite{sabetzadeh2019majority} and edge detection~\cite{strollo2020comparison} are used to evaluate the practical performance of the approximate multipliers generated by OpenACM. In image blending, an 8-bit unsigned multiplier processes two grayscale images pixel by pixel, with results scaled back to 8 bits. For edge detection, convolution and squaring employ a 16-bit signed approximate multiplier, while the square root is computed exactly. Image quality is measured by peak signal-to-noise ratio (PSNR), using the exact multiplier as the baseline. Typically, PSNR below 30 dB indicates visible degradation, whereas values above 40 dB imply near-identical quality.

As shown in \autoref{tab:Image Processing}, Appro4-2 achieves high accuracy and can effectively replace exact multipliers in image processing. Log-our further improves accuracy over the LM~\cite{mitchell2009computer}. In image blending, LM produces unsatisfactory results with PSNR generally below 30 dB, while Log-our consistently surpasses this threshold, making it viable for energy-efficient applications. In edge detection, LM reaches about 30 dB, whereas Log-our achieves 45 dB, delivering quality comparable to the exact design. Overall, the proposed OpenACM architecture supports both accuracy-critical and energy-constrained scenarios, enabling flexible circuit selection tailored to diverse application needs.

\begin{table}[tbp]
\centering
\caption{Influence of approximate multiplier on Top-1 and Top-5 scores for the ResNet-18 network.}
\label{tab:neural networks}
\begin{tabular}{ccccc}
\toprule
Multiplier Type & Top-1 & Top-5 &NMED & MRED \\
\midrule
Exact & 0.677 & 0.873 & - & - \\
Appro4-2 & 0.668 & 0.880 & 1.70E-09 & 1.27E-10\\
Log-our & \textbf{0.680} & \textbf{0.870} & 4.40E-03 & 1.55E-02\\
LM~\cite{mitchell2009computer} & 0.610 & 0.842 & 2.79E-02 & 9.44E-02\\
\bottomrule
\end{tabular}
\end{table}

Neural networks are widely applied in computer vision and image classification, where approximate multipliers can often replace exact ones during forward inference. We utilized the pre-trained ResNet-18~\cite{he2016deep} model on the ILSVRC2012 dataset as a baseline and tested various approximate multipliers for the Top-1 and Top-5 accuracies of this neural network. The floating-point weights and inputs are quantized to 32-bit fixed-point for computations. In addition, the normalized mean error distance (NMED) and mean relative error distance (MRED) are calculated as error metrics of the approximate multipliers for reference. 

As shown in \autoref{tab:neural networks}, both Appro4-2 and Log-our reduce power consumption significantly without degrading classification accuracy. Specifically, Appro4-2 achieves 17\% power savings, while Log-our achieves 64\% compared to the exact multiplier. Thanks to the inherent error resilience of deep neural networks, such reductions can be realized with no loss of inference quality. Notably, although Appro4-2 introduces smaller absolute errors, its one-sided distribution leads to systematic deviations in accumulated results. In contrast, Log-our produces bidirectional errors resembling zero-mean perturbations, which act as noise regularization and enhance generalization. Consequently, Log-our attains slightly higher Top-1 accuracy than the exact multiplier, while Appro4-2 shows only minor Top-5 improvement. The conventional LM suffers from large errors, causing a severe accuracy drop, further validating the effectiveness of the proposed Log-our.

\subsection{IS-Based Yield Analysis}

\autoref{tab:yield111} presents a comparison of yield estimation between Monte Carlo (MC) and Mean-shifted Importance Sampling (MNIS)~\cite{dolecek2008breaking}. 
To further accelerate circuit-level simulations, we use trimmed SRAM arrays with only two columns (N$\times$2), while retaining the full WL parasitics of the original arrays. This preserves the WL RC loading seen by the drivers but reduces the number of simulated bitline columns, yielding substantial runtime savings.
The figure of merit (FoM) shown in the table is defined as $\mathrm{std}(P_f) / P_f$, and $\mathrm{std}(P_f)$ represents the standard deviation of the estimated failure rate $P_f$. As shown, MC requires a substantially larger number of simulations (e.g., 41,500 for the 64$\times$2 case), whereas MNIS achieves comparable accuracy with significantly fewer runs (only 4,260), yielding a speedup of approximately 9.7$\times$. Similar trends are observed for smaller circuits, where MNIS achieves about 18$\times$ speedup for the 16$\times$2 case and 10$\times$ for the 32$\times$2 case. Furthermore, as the circuit size increases, MNIS continues to exhibit superior scalability and robustness, thereby reinforcing its effectiveness. Overall, these results highlight the advantages of importance sampling-based approaches in yield estimation and demonstrate their strong potential to replace conventional MC methods in large-scale circuit characterization.
 
\label{sec:yield}
\begin{table}[tb]
\centering
\caption{Comparison of MC and MNIS Yield Analysis Methods on Various SRAM Sizes.}
\label{tab:yield111}
\resizebox{\linewidth}{!}{%
\begin{tabular}{cccccccc}
\toprule
\multirow{2}{*}{Metric}& \multicolumn{3}{c}{Monte Carlo} & \multicolumn{3}{c}{MNIS~\cite{dolecek2008breaking}} & \multirow{2}{*}{Speedup} \\ \cmidrule(r){2-4} \cmidrule(r){5-7}
& $P_f$ & FoM & \#Sim. & $P_f$ & FoM & \#Sim. \\ \midrule
16 $\times$ 2 & 1.6E-4  & 0.1 & 55,600 & 3.2E-4 & 0.05 & 2,985 & 18$\times$\cellcolor{green!15} \\
32 $\times$ 2 & 6.4E-2 & 0.17 & 22,900 & 1.7E-2 & 0.15 & 2,260 & 10$\times$\cellcolor{green!15} \\
64 $\times$ 2 & 3.9E-3 & 0.05 & 41,500 & 1.5E-3 & 0.03 & 4,260 & 9.7$\times$\cellcolor{green!15} \\ \bottomrule
\end{tabular}%
}
\end{table}

\section{Conclusion}
\label{sec:conclusion}
In this work, we presented OpenACM, the first open-source, accuracy-aware compiler for SRAM-based approximate DCiM architectures. While OpenACM provides a solid foundation, several key extensions remain. Our near-term priorities include completing the automated layout generator for custom SRAM macros to enable a fully physical-aware flow, and extending the front-end and multiplier library to support native floating-point operations. We also plan to develop an automated DSE engine to jointly optimize multiplier choice, precision, and array organization for application-specific efficiency. We believe OpenACM will continue to support and accelerate research in energy-efficient computing.

\bibliographystyle{IEEEtran}
\bibliography{refs}

\end{document}